\documentclass[12pt]{article}

\newcommand{\skyp}[1]{}

\newcommand{\be}{\begin{equation}}
\newcommand{\ee}{\end{equation}}
\newcommand{\bea}{\begin{eqnarray}}
\newcommand{\ra}{\rightarrow}
\newcommand{\eea}{\end{eqnarray}}
\newcommand{\N}{{\cal N}}

\def\ll{{<\!\!<}}

\font\mybb=msbm10 at 12pt

\def\bb#1{\hbox{\mybb#1}}

\def\Z {\bb{Z}}

\def\id{\protect{{1 \kern-.28em {\rm l}}}}
\def\ttau{{\tilde \tau}}

\makeatletter
\renewcommand\section{\@startsection {section}{1}{\z@}%
                                   {-3.5ex \@plus -1ex \@minus -.2ex}%
                                   {2.3ex \@plus.2ex}%
                                   {\normalfont\large\bfseries}}
\renewcommand\subsection{\@startsection{subsection}{2}{\z@}%
                                   {-3.25ex\@plus -1ex \@minus -.2ex}%
                                   {1.5ex \@plus .2ex}%
                                   {\normalfont\normalsize\bfseries}}
\makeatother

\newcommand{\sect}[1]{\section{#1}\setcounter{equation}{0}}

\begin{document}

\begin{titlepage}

\bigskip
\hfill\vbox{\baselineskip12pt
\vbox{\hbox{UCLA/03/TEP/18}}}
\bigskip\bigskip

\centerline{\Large ${\cal N} = 1^*$ in 5 dimensions: Dijkgraaf-Vafa 
meets Polchinski-Strassler}

\bigskip
\bigskip
\centerline{ Iosif Bena,$^1$ Radu Roiban$^2$}
\bigskip
\medskip
\centerline{$^1$\it Department of Physics and Astronomy}
\centerline{\it University of California, Los Angeles, CA\ \ 90095}
\centerline{ iosif@physics.ucla.edu}
\bigskip
\centerline{$^2$\it Department of Physics}
\centerline{\it University of California, Santa Barbara, CA \ \ 93106}
\centerline{ radu@vulcan2.physics.ucsb.edu}
\bigskip
\bigskip

\begin{abstract}

One of the powerful techniques to analyze the 5 dimensional Super
Yang Mills theory with a  massive hypermultiplet (${\cal N}=1^*$) 
is provided by the AdS/CFT correspondence. It predicts that,
for certain special values of the hypermultiplet mass,
this theory develops nonperturbative branches of the moduli space
as well as new light degrees of freedom.

We use the higher dimensional generalization of the matrix model/gauge
theory correspondence and recover all the prediction of the
supergravity analysis. We construct the map between the four
dimensional holomorphic superpotential and the five dimensional action
and explicitly show that the superpotential is flat along the
nonperturbative branches. This is the
first instance in which the Dijkgraaf-Vafa method is used to 
analyze intrinsically higher dimensional phenomena.

\end{abstract}

\end{titlepage}

\newpage

\baselineskip=18pt

\sect{Introduction}

The five dimensional supersymmetric Yang Mills coupled with one
massless  
hypermultiplet has 16 supercharges and is the low energy theory 
living on a stack of D4 branes. Due to its nonrenormalizability 
this theory is perturbatively ill-defined. However, its phase structure 
is more complicated and the high energy regime reveals a six
dimensional nature. In a certain energy range 
the generalization \cite{imsy} of the $AdS/CFT$ correspondence implies
that this theory is dual to supergravity on the near horizon geometry 
of D4 branes.

Deforming the brane theory by a mass term for the hypermultiplet 
(and thus obtaining the five dimensional ${\cal N}=1^*$ theory)
corresponds on the gravity side to turning on 
non-normalizable 
modes of the Ramond-Ramond (RR) 2 form and the Neveu-Schwarz (NSNS) 3
form field strengths. In 
\cite{rc} it was realized that this perturbation can be interpreted
as a Melvin twist in the compactification of M5 branes to D4 branes. 
This observation led to  the construction of the exact supergravity 
dual of the mass-deformed superYang-Mills theory in five dimensions.

The supergravity dual has a rather odd feature: for specific values of 
the five dimensional gauge coupling (such that 
$A={\textstyle\frac{1}{4\pi^2}}g_{5}^2 N m_0$ is integer), the D4 branes
can polarize \cite{myers} 
into $A$ circular NS5 branes, and the polarization radii
are moduli. This implies that for these values of the coupling
constant the theory has  a ``polarization'' branch, which intersects
the Coulomb branch at points with enhanced gauge symmetry. 
All  vacua on the polarization branch reduce in the classical limit
to  a {\em single} vacuum.

We should stress that the existence of polarization branches only at
very specific values of the parameters is a stringy phenomenon. 
Indeed, if ${\textstyle\frac{1}{4\pi^2}}g_{5}^2 N m_0$ were
not an integer, one still finds a solution to the supergravity
equations of motion. However, this solution has fractional NS5
brane charge and should be discarded since it cannot be a solution of 
string theory.

To gain an intuitive understanding of
the quantum nature of this branch it is perhaps
instructive  to recall the 4-dimensional $\N=1^*$ theory \cite{n=1*},
to which the  5-dimensional $\N=1^*$ theory is related by naive
dimensional 
reduction and mass deformation. In that case, to one
classical vacuum with unbroken gauge symmetry correspond $N$ quantum
vacua, distinguished by the phase of the gaugino condensate. This
phase can be interpreted as a (discrete) modulus. In the case we are
interested in, to one 
classical vacuum with unbroken gauge symmetry corresponds a continuum 
of quantum vacua, distinguished by the polarization radius (which in
field theory language represents the expectation value of 
an operator bilinear in fields).

The purpose of this paper is to recover the predictions of the string 
dual to the five dimensional $\N=1^*$ theory using the extension of
the matrix  model -- gauge theory relation to theories in more than four
dimensions \cite{dv4}.
The original formulation of this relation 
\cite{dv123} allows one to compute 
holomorphic quantities of a 4 dimensional gauge
theory using a matrix model. In the spirit of deconstruction,
one can 
interpret a $4+d$ dimensional theory as a 4-dimensional one
with an infinite set of fields labeled by the remaining $d$
coordinates. After a suitable supersymmetry breaking deformation, which
lifts all but a discrete set of points of the moduli space,
it is possible to
use the original gauge theory/matrix model relation 
to compute the holomorphic superpotential  of  this ``effective''
4-dimensional theory. Because of the remaining coordinates labeling the
fields, the matrix model becomes a $d$ dimensional euclidean
bosonic matrix field theory.

In the particular case we are interested in, this is a matrix quantum
mechanics. Once its action is found, it is still necessary to solve it
and find the five dimensional interpretation of the 
four dimensional holomorphic  quantities which can be computed.

A particular deformation preserving four supercharges
of the  ${\cal N}=1^*$ theory compactified on a circle 
was recently studied in
\cite{holo}, where the matrix quantum mechanics 
corresponding to this theory was identified and solved. We will therefore
need only to extract the five dimensional physics from the four 
dimensional ${\cal N}=1$ holomorphic quantities. As we will see later,
the fact that the solution is not known for a generic deformation 
will not be a problem for identifying the features predicted by
the supergravity analysis.

We will proceed in \S2 to review the supergravity dual of the 5
dimensional  ${\cal N}=1^*$ theory and describe some of its
properties. In \S3 we will review the DV description \cite{holo} of the
compactified version of this theory. 
We will then proceed in \S4 to discuss the features of the
superpotential and then match them with the supergravity predictions
in \S5 and \S6.

Various discussions on the five dimensional ${\cal N}=1^*$ theory have
appeared  in the literature. In particular, a solution for the 
theory with an $SU(2)$ gauge group compactified on a circle 
was proposed by Nekrasov \cite{nekrasov}. It turns out that the
resulting superpotential agrees, up to field redefinitions, with the
one computed in \cite{holo} using DV techniques. The solution 
proposed in \cite{nekrasov} exhibits a  symmetry under certain shifts of
the mass of the hypermultiplet. As a byproduct of the supergravity 
discussion in \S2, we will see that the extension of this symmetry to
an $SU(N)$ gauge group corresponds to a symmetry of the dual geometry.

\sect{The theory and its supergravity dual}

As it is well known, the gauge-gravity duality \cite{malda} relates
field theory living on a large number of D-branes to string theory in
their near horizon geometry. Our discussion
focuses on a mass deformation of the field theory living on a set of $N$
coincident D4 branes. 

The original theory has 16 supercharges and its field content is a
5-dimensional vector 
multiplet and a hypermultiplet. The five real scalars 
correspond to 
the five directions transverse to the branes, and transform in the 
vector representation of the $SO(5)$ R-symmetry group. 

The duality between this  maximally supersymmetric gauge theory and
string theory  was first investigated in \cite{imsy}. This
theory is not conformally invariant, and thus different energy ranges have
different weakly coupled descriptions. Since the coupling constant has
positive dimension, the field theory is (generically) weakly coupled in
the infrared. As one increases the energy, the type IIA supergravity
background becomes weakly curved and thus provides an appropriate description
of the theory. At even higher energies, the dilaton becomes  
large enough for the M-theory circle to decompactify and for the 11
dimensional supergravity to become the appropriate description of the
theory.

This duality can be perturbed on both sides. On the field theory
side, we introduce a mass term for the hypermultiplet and obtain the five
dimensional ${\cal N}=1^*$ theory. This theory preserves 
8 supercharges and has a Coulomb branch parameterized by the
expectation values of operators built out of the vector
multiplet scalar. On the supergravity side this mass deformation
corresponds 
to turning on a non-normalizable mode \cite{bdhm-bklt}  
of the RR two form and NS-NS three form field strengths in
directions transverse to the branes. This deformation removes the
freedom of distributing the branes in four of the five transverse
directions, which is the supergravity manifestation of four of the
five original Coulomb branch directions being lifted.

Let us now turn to a more detailed description of the supergravity
solution. As usual, this description of the theory is valid 
as long as the string coupling is small and the background geometry is
weakly curved. This raises the question of interpreting the various
singularities that the solution might develop. We will show that the
singularities appearing in the weak coupling regime are physical and
have a standard field theoretic interpretation.

\subsection{The expected - singularities in the dual geometry, and
their physical meaning.}

The supergravity background dual to the 5 dimensional ${\cal
  N}=1^*$ theory was investigated in \cite{rc}, using techniques
  similar to those employed by Polchinski and Strassler \cite{ps} in
  the analysis of the 4 dimensional ${\cal N}=1^*$ theory. It  
was also observed that this background can be obtained by
  longitudinally reducing the 11 dimensional supergravity  
background sourced by M5 branes with ``Melvin twists'' on two angular
directions. In the absence of the M5 branes, this reduction yields the
supersymmetric flux-5-brane of Gutperle and Strominger \cite{gs}. Adding
the M5 branes gives a type IIA solution which is a superposition of the
  flux 5-brane and D4 branes\footnote{Similarly to the original
  flux-brane, this solution preserves 8 supersymmetries
  \cite{fig}.}. The relation between the gauge theory mass parameter
and supergravity twist can be easily established by a first order
computation. It turns out (perhaps not surprisingly) that the mass of
the ${\cal N}=1^*$ hypermultiplet is proportional to the Melvin twist
\begin{equation}
m= B.
\label{mass}
\end{equation}
Clearly, in the limit of vanishing mass (vanishing twist) one recovers
the D4 brane solution and the corresponding maximally supersymmetric
gauge theory.  

The exact background dual to the 5 dimensional $N=1^*$ theory is:
\bea
g_s ^{-4/3} e^{4\phi/3}&=& Z^{-1/3}+B^2 \, Z^{2/3}(\rho_1^2
+\rho_2^2) \equiv \Lambda \nonumber\\
ds_{10}^2 &=& \Lambda^{1/2}( Z^{-1/3} dx_{\parallel}^2+
Z^{2/3} dx_{\perp}^2)-B^2 \Lambda^{-1/2}Z^{4/3}(\rho_1^2  d 
\phi_1+\rho_2^2  d  \phi_2)^2~~,\nonumber\\
g_s C_{ \phi_1}&=& \Lambda^{-1} \rho_1^2 B Z^{2/3}~~~, ~~~~~~~
g_s C_{ \phi_2}= \Lambda^{-1} \rho_2^2 B Z^{2/3} ~~,\nonumber\\
g_s F_4&=& *_5 d Z^{-1} ~~, ~~~~~~~
*_5\! H_3= B (\rho_1^2 d \phi_1+\rho_2^2 d \phi_2  )\wedge d Z
\label{bg}
\eea
where $\rho_1,\phi_1$ and $\rho_2,\phi_2$ are the radial and angular
coordinates parameterizing the two transverse 2-planes where the
Melvin twists were made, $x_7$ is the direction  corresponding to the
scalar in the vector multiplet and $*_5$ is the Hodge dual taken with
the flat metric in the directions transverse to the branes. 

It is important to note that a D4 
brane displaced in the $(\rho_1,\phi_1)$ or $(\rho_2,\phi_2)$ plane
feels a potential forcing it toward the origin. This potential only 
vanishes when the Melvin twists vanish,
which reflects the fact that in this limit maximal supersymmetry is 
restored.

The solution (\ref{bg}) is not weakly curved everywhere. For example,
when all the branes are coincident, the Ricci scalar becomes large in
a neighborhood of size $g_5^2 N B^2$ around the origin. The situation 
does not improve substantially even when the branes are distributed on
the Coulomb branch with some density $\rho(y)$. The harmonic function 
changes to 
\be
Z=\int {\rho(y)dy \over (\rho_1^2+\rho_2^2+(x_7-y)^2)^{3/2}}
\ee
which generically leads to a divergent Ricci scalar at
$\rho_1=\rho_2=0$. Even though these singularities come with a zero
size horizon (as it is easy to see from the metric (\ref{bg})), it is
important to understand whether they are physically meaningful.

A similar singularity exists at the location of the D4 branes
in the undeformed solution. In that case the
singularity is interpreted as a change in the correct description of
the theory: from supergravity away from the branes to the 
perturbative Yang-Mills theory in 5 dimensions close to them.
Given its similarity to the D4 brane case, it is reasonable to expect
that the singularity of (\ref{bg}) is
physical and signifies the presence of a weakly coupled field theory
description. 

Two criteria help us decide. According to one of them  
\cite{maldacena}, a singularity is physical if the $g_{00}$ component
of the metric does not diverge as one
approaches it. As one can easily see
from (\ref{bg}), our singularity 
clearly passes this test. The other criterion  
\cite{guster} implies that a singularity of a supergravity solution
has a field 
theoretic interpretation  if it is possible to construct a family of
solutions 
with the singularity covered by horizons of finite size, which has the
original solution as a smooth limit. 

To check whether our solution passes the second test, we should attempt 
to make it near extremal. It turns out that 
this is not very difficult: using the relation between (\ref{bg}) and
M5 branes, it is easy to see that the near extremal background is simply
the KK reduction of the near-extremal M5 brane background. The dilaton is
not modified, and the string frame metric is simply
\begin{eqnarray}
ds_{10}^2 &=& \Lambda^{1/2}\left[ Z^{-1/3} f(r) dt^2 + Z^{-1/3}
  dx_{_{1,2,3,4}}^2+ 
Z^{2/3} \left( \frac{dr^2}{f(r)} + r^2  d
\Omega_4^2\right)\right]\nonumber\\ 
&&-B^2 \Lambda^{-1/2}Z^{4/3}(\rho_1^2
d\phi_1+\rho_2^2 d\phi_2)^2,
\label{bh}
\end{eqnarray}
where $f(r)=1-{R_{H}^3 \over r^3}$, $R_H$ is the horizon area
and $r^2=x_7^2+ \rho_1^2 + \rho_2^2 $. It is a simple
exercise to compute the Einstein frame horizon area of (\ref{bh}), and 
to see that it vanishes in the zero temperature limit. 

Before we proceed let us remark a rather odd property of the
theory implied by the nonextremal solution (\ref{bh}): 
\begin{equation}
S=\frac{2\pi^2}{3}R_H^4 Z(R_H)^{1/2}~~~~~~~~~~~
T_H=\frac{1}{3\sqrt{3}\pi R_H Z(R_H)^{1/2} }~.
\label{thermal}
\end{equation}
Thus, for a fixed temperature, the horizon area is
independent of the
Melvin twist\footnote{The equation (\ref{thermal}) implies that this
holds for all distributions which are twist-independent. It is however not
hard to see that the only finite distribution for which (\ref{bh}) is a 
solution is the one with all the branes at
the origin of the Coulomb branch.}. 
This implies that the entropy of the dual field theory is independent
of the  hypermultiplet mass.  
Though such a behavior might appear strange, it is not unique to this
system. A similar phenomenon occurs for a field compactified on a circle
with twisted boundary conditions $\psi(0)=\psi(2 \pi) e^{i \alpha}$. 
As one shifts $\alpha$ by $2 \pi$ the massless states become massive,
while some  massive states become massless. However $\alpha \ra
\alpha+2 \pi$ is a symmetry of the spectrum and thus the entropy is
invariant under this transformation. We will further  discuss  the gauge
theory implications of this observation in \S5.

\subsection{The Unexpected - Brane Polarization}

Apart from finding the supergravity solutions dual to a generic point on
the Coulomb 
branch of the 5-dimensional ${\cal N}=1^*$ theory, in \cite{rc} it was
observed (using methods  
similar to those of \cite{ps})  that for very specific
values of the hypermultiplet mass (such that
$\frac{1}{4\pi^2}g_{5}^2 m N\in \Z$) the D4 branes can polarize 
into a circular NS5 brane and moreover, that the radius of polarization
is a modulus. For other values of $m$ (such that
$\frac{1}{4\pi^2}g_{5}^2 m N'\in \Z$ with $N'$ being a divisor of $N$), 
it is possible for $N'$ of the D4 branes to polarize into NS5 branes,
while the  other $(N-N')$ branes are free to either move on the Coulomb
branch or  polarize into more NS5 branes.  Thus, if ${4\pi^2 \over
g_5^2 m} \notin \Z$, there is no 
polarization branch while, if ${4\pi^2 \over g_5^2 m}\in \Z $, the
polarization branch intersects the Coulomb branch in points with
nonabelian symmetry $SU({4\pi^2 \over g_5^2 m})$.

This puzzling phenomenon can be explained rather easily by recalling 
the M-theory origin of the background. Let us pick a uniform, circular
array of $N$ M5 branes in one of the $(\rho, \phi)$ planes.
If the Melvin twist accompanying the reduction to 10 dimensions matches
the ends of two neighboring branes, the resulting type IIA
configuration will be one circular NS5 brane with $N$ units of D4
brane charge. If however the twist does not match brane ends, the
descending configuration will not have integer local NS5 charge, and
will not be a consistent solution of string theory.
In general, only shifts which match brane ends yield consistent
IIA solutions with D4 branes inside cylindrical NS5 branes. 

The Killing vector along $x_{11}$ is proportional to the 
Killing vector along $\phi$. This implies that the matching of M5 
brane ends is independent of the radius. Thus, the D4-NS5 
configuration is a solution at any radius, and therefore the 
potential for the radius of the brane configuration is flat.
This reproduces the  result of the  Polchinski-Strassler-type
analysis \cite{rc}.

To summarize, supergravity predicts a dramatic change in the infrared
physics depending on the integrality properties of 
$A=\frac{1}{4\pi^2}g_{5}^2 m N$.  When the branes are coincident,
the  solution has a ``good'' singularity in the infrared, 
signifying the flow to a weakly coupled gauge theory at low
energies. When $A$ is not an integer, one can soften the singularity by 
distributing the branes on the Coulomb branch, but the singularity never
disappears. This implies that infrared physics of these ${\cal N}=1^*$
theories has a weakly coupled  field theory description.

When $A$ is integer one can also soften the singularity by 
going on the polarization branch.  The geometry is still given by
(\ref{bg}),  where now the harmonic function $Z$ is sourced by branes
on a circle. At a generic point on this branch 
supergravity is weakly coupled everywhere in the bulk\footnote{This
is easy to see, because on the polarization
branch the brane configuration is locally an NS5 brane.}. 
This implies that for integer $A$ and nonvanishing expectation 
value of the bilinear
operator constructed out of the two complex scalars in the
hypermultiplet, the infrared theory remains strongly coupled. 

The supergravity solution discussed above predicts that 
further interesting
phenomena occur in the theories with strong infrared dynamics; 
we will return to them in \S6.

\subsection{On a conjectured symmetry of the gauge theory}

As briefly mentioned in the introduction, the construction of 
the supergravity dual of the five dimensional ${\cal N}=1^*$
theory allows us to prove the conjecture
\cite{nekrasov} that this theory is invariant under shifts 
of the mass parameter $m$ by a
quantity proportional to the inverse of the five dimensional 
gauge coupling. It is not hard to translate the precise form of the
symmetry transformation in supergravity language using the
observation that the mass of the hypermultiplet is 
proportional to the Melvin twist relating the 10 and 
11 dimensional geometries (\ref{mass}). 
The shift of the mass parameter conjectured in
\cite{nekrasov} to be a symmetry of the theory 
corresponds to changing the Melvin twist by $2 \pi$.

A similar invariance in the case of an exactly solvable string theory
does exist, as discussed in \cite{rt}. By explicitly computing the
spectrum it was shown that
$2 \pi$ shifts of the Melvin twist are nontrivial automorphisms both 
of the Kaluza-Klein tower of supergravity states and of the
full string spectrum. It is however difficult to make similar
statements about string interactions. 

Even though it is not possible to directly extend even the simple
spectrum argument to the case we are interested in\footnote{since 
we do not have a
quantum realization of M-theory}, it is possible however to reach the
same conclusion as in the string theory by using symmetry arguments. 
The basic observation is that the
supergravity fields and their Kaluza-Klein modes form short
representations of the supersymmetry algebra, while the massive
M-theory modes form long representations. Moreover, $2 \pi$ shifts of 
the Melvin twist leave invariant the 11 dimensional supergravity spectrum 
and the boundary conditions defining the reduction to 10 dimensions. 
Therefore, these shifts are automorphisms of
the 10 dimensional spectrum of supergravity and KK
states. 

These ingredients are sufficient to establish the 
shift invariance at the level of holomorphic quantities. 
From the invariance of the background it follows 
that the 1-point functions (which are also insensitive to bulk
interactions) of operators forming short multiplets are unchanged (up
to field redefinitions) by $2 \pi $ shifts of the Melvin twist. 
Standard arguments imply that these quantities  determine
the superpotential via
\begin{equation}
\langle O_n\rangle = \frac{\delta W_{\it eff}}{\delta
g_n}~~~~~~~~~~~~~~ W_{\it tree}=\sum g_n O_n~~,
\end{equation}
up to $g_n$-independent terms.
Therefore, the superpotentials of theories related by $2 \pi $ shifts
of the mass parameter of the hypermultiplet are the same up to field
redefinitions, which proves the conjecture of \cite{nekrasov}.

It is interesting to note that the shift symmetry of the 5 dimensional
$\N=1^*$ theory holds also at the Kahler potential level. Indeed, due
to the extended supersymmetry of the field theory, the Kahler
potential and the superpotential are related. Proving this statement
from a bulk standpoint is a probably a hard problem given our poor
understanding of the quantum M theory. However, this is not unexpected, 
since in the flat space example discussed in \cite{rt},
the full string spectrum is symmetric under these shifts.

\sect{Matrix quantum mechanics; the 4d superpotential}

As we have briefly described in the introduction, the relation between
matrix models and 4 dimensional gauge theories can be extended to 
a relation between $d$ dimensional Euclidean bosonic field theories and 
$4+d$ dimensional gauge theories. 

To do this one formally separates the $4+d$ coordinates into $4$
``external'' coordinates $x$ and $d$ ``internal''
coordinates $y$, and treats each field $\Phi(x,y)$, as an
infinite set of 4 dimensional fields labeled by $y$. The kinetic 
terms along the $y$ coordinates appear as (super)potential terms in
the 4 dimensional theory. This deconstruction procedure
\cite{siegel,arkani-hamed} can be carried out either at the
component level or in superspace and yields a 4
dimensional (super)potential presented as an integral over the $d$
internal coordinates. Quite generally, the resulting 4 dimensional theory has
extended supersymmetry. Adding 
suitable supersymmetry breaking terms it is then possible, using
the gauge theory/matrix model relation \cite{dv123}, to compute the
holomorphic superpotential in the resulting ${\cal N}=1$ theory. The
matrix model obtained in this way has an infinite 
number of fields labeled by the $d$ internal coordinates; it is in
fact a $d$-dimensional Euclidean matrix field theory.

This construction leads to the idea that every $d$
dimensional bosonic matrix field theory describes the sector of some
$4+d$ dimensional supersymmetric gauge theory which becomes
the holomorphic sector when ``deconstructed'' to 4 dimensions.
However, not all $4+d$ dimensional theories described in this way have
$4+d$ dimensional Lorentz invariance. Indeed, it is not hard to see
that only superpotentials which are linear in derivatives along the
internal directions give regular kinetic terms in $4+d$ dimensions.

In this section we are interested in slightly extending the analysis 
\cite{holo} of the 5 dimensional ${\cal N}=1^*$ theory compactified 
on a circle of radius $\beta$. 
According to the general philosophy described above we should solve a
matrix quantum mechanics. It turns out however that the gauge
symmetry can be used to set one of the three fields to a constant,
makings things somewhat simpler.

To understand the simplification of the matrix quantum mechanics
associated to the ${\cal N}=1^*$ theory let us first review the case
of pure 5-dimensional super-Yang-Mills theory. By compactifying it on
a circle the component of the gauge field along the compact direction,
$A_t$, becomes a scalar, and combines with the scalar of the vector
multiplet of the five dimensional theory to form the complex scalar
$\Phi_3$.  Using local gauge transformations, one
can set the nonconstant part of $A_t$ to zero. However, one still has
large gauge transformations which shift the eigenvalues of $A_t$ by
$2\pi /\beta$. Thus, the natural gauge invariant quantity is
not $A_t$ but the holonomy of the gauge field  along the compact
direction, $e^{i \beta A_t}$, which is independent of $t$. 
Furthermore, due to ${\cal N}=1$ supersymmetry in 4 dimensions
the actual variable is the ``holonomy'' of the superfield $\Phi_3$.
Indeed, since the gauge parameter is a chiral superfield, one can use
it to set the whole superfield $\Phi_3$ to a constant, up to shifts by
$2\pi i/\beta$.
The net result of this gauge choice is that the dynamics of
$\Phi_3$ is 
captured by  $U \equiv e^{\beta \Phi_3}$. As discussed in
\cite{dv4}, this implies that in the case of pure 5-dimensional ${\cal
N}=1$ gauge theory, the matrix quantum mechanics reduces to a matrix
model.

In the case of 5 dimensional ${\cal N}=1^*$ theory things are slightly
more complicated due to the presence of the hypermultiplet whose 
$t$-dependence cannot be gauged away. Thus, one has to study the 
full matrix quantum mechanics which  nevertheless has one constant
field $\Phi_3$.
Its action is given by
\be
W[\Phi_i]= {\rm Tr}[i \Phi_1 D \Phi_2
 + m \Phi_1 \Phi_2 ]
\label{w2}
\ee
where $D \Phi_2 = \partial_t \Phi_2 + [\Phi_3,\Phi_2]$, and $\Phi_3$ is
$t$-independent.

This superpotential preserves ${\cal N}=2$ in 4 dimensions. To use the 
Dijkgraaf-Vafa approach \cite{dv123} it is necessary to deform it by an
${\cal N}=2 \rightarrow {\cal N}=1$ breaking 
term\footnote{This procedure, which depending on the deformation
selects a point of the ${\cal N}=2$ Coulomb branch as the ${\cal N}=1$
vacuum, was used to derive the known Seiberg-Witten solutions of
ordinary four dimensional ${\cal N}=2$ theories.}. In principle, any
function of $\Phi_3$ which is invariant under $\Phi_3 \rightarrow
\Phi_3 + 2 \pi i/\beta$  (i.e. any function of $U$) is a suitable
deformation. Furthermore, each such function describes a point on
the 4-dimensional ${\cal N}=2$ Coulomb branch. Thus, to cover the 
whole moduli space
it would be necessary to use a function with a ${\cal O}(N^2)$
arbitrary coefficients. This seems a formidable task, especially
because we are interested in finding the precise form of the effective
superpotential. 
We will restrict to the theory studied in \cite{holo} --  
the compactified ${\cal N}=1^*$  deformed by the tree level
superpotential $\mu\cosh(\beta \Phi_3)$ -- and show that
the accessible subspaces of the moduli space already exhibit
the nontrivial behavior described in the previous section.

The
relevant matrix quantum mechanics was solved using techniques
similar to those used to study the Leigh-Strassler deformation of the
4 dimensional $N=4$ Super Yang Mills theory \cite{leigh,dorey}.
The starting point is the partition function:
\be
Z=e^F=\int \prod_{1,2,3}[d \Phi_i] \exp(-{1 \over \beta g_s} \int dt
W[\Phi_i])
\ee
where 
\be
W[\Phi_i]= {\rm Tr}[i \Phi_1 D \Phi_2 + m \Phi_1 \Phi_2 
+ \mu (\cosh(\beta \Phi_3)-1)]
\label{breaking}
\ee
with $D \Phi_2 = \partial_t \Phi_2 + [\Phi_3,\Phi_2]$. As usual, the
effective 
superpotential is constructed out of the derivatives of the free
energy $F$.

One first integrates out the fields $\Phi_1$ and $\Phi_2$, and obtains
a potential which  depends only on the holonomy matrix $U$. Due to its
coordinate independence this reduces again the matrix quantum
mechanics to a matrix model. After a few rather technical steps
\cite{holo}, one finds: 
\bea
W_{\it eff} &=& N {\partial F \over \partial S} - 2 \pi i \tau S 
~~~~~~~~
\tau = {\theta\over 2\pi}+{2\pi i\over g_4^2}\\ 
2 \pi i S &=& {d h(t) \over d t}= h'(t)~, ~~~~~~~~   
{\partial F \over \partial S} = t h'(t) - h(t)~, ~~~~{\rm where} \\
h(t) &=& {\mu \over \sin  \beta m /2} 
{\theta_1(\beta m /2|t) \over \theta'_1(0|t)}~~,
\label{ww}
\eea
and $t$ is the modular parameter of an auxiliary Riemann surfaces
appearing in the solution of the $U$ matrix model.
This gives implicitly the superpotential in terms of the glueball 
superfield $S$, as well as the effective coupling constant of the 
remaining $U(1)$ vector multiplet:
\be
2 \pi i\,\, \tau_{\it eff} =  {\partial^2 F \over \partial S^2} 
~~.
\ee

Knowledge of the effective superpotential leads to knowledge of vacua
of the theory. In the limit in which supersymmetry is restored they 
become vacua of the resulting ${\cal N}=2$ theory.
Minimizing $W_{eff}$ with respect to $t$, we obtain
\be
h''(t) (t N - \tau) =0
\ee
which has 2 classes of solutions:
\be
t = {\tau \over N}  ~~~~~~~~  {\rm and} ~~~~~~~~  h''(t)=0
\label{vacua}
\ee

The distinction between the two types of solutions is crucial. While
in the first case the relation between $S$ and $t$ is well-defined,
at the points where $h'=0$, this relation becomes
degenerate. As we will see shortly, the effective coupling $\tau_{\it
eff}$ is not well defined at those points. This suggests a breakdown
of the effective description, perhaps accompanied by the appearance of
new light degrees of freedom. We will come back to this phenomenon in
\S6 and link it with similar features exhibited by the
supergravity solution.

Let us consider for now only the ``regular'' solution $t = {\tau \over
N}$. Since a nonvanishing $\theta$ angle breaks 5-dimensional Lorentz 
invariance, we will set it to zero. The vacuum
value of the effective superpotential is then  
\be
W_{eff} = - {N\mu \over \sin ~ {\beta m \over 2}} {\theta_1({\beta m
\over2}|{2 \pi i \over g^2_4
N} ) \over \theta'_1(0|{2 \pi i \over g^2_4 N}) } = - {N\mu \over \sin ~
{\beta m \over 2}} 
{\theta_1 \left({\beta m \over 2 }|{2 \pi i \beta \over g_5^2 N}
\right) \over \theta'_1\left(0|{2 \pi i
\beta \over g_5^2 N} \right)}.
\label{Wmin}
\ee

Before proceeding to analyzing in detail the physics of this
superpotential it is important to point out that the real part of the
mass parameter $m$ in (\ref{w2}) corresponds to a Lorentz-breaking 
term in the  5 dimensional Lagrangian,
and therefore does not interest us here. The imaginary part of $m$ is
the mass parameter of the 5-dimensional  hypermultiplet: $m= i m_0$. 
To fix notation, let us state that the 5 dimensional Yang-Mills
coupling constant is related to the 4-dimensional one by: $g_5^2 \equiv
g_{YM-5}^2 = g_4^2 \beta$, where $\beta$ is the length of the
compactified dimension.

In the next section we will analyze the effective superpotential
(\ref{Wmin}) and extract the information relevant to the 5 dimensional
${\cal N}=1^*$ theory. For this purpose we need to examine the limit 
$\beta \rightarrow \infty$ keeping $g_5^2$ fixed. Since in this limit
the four dimensional coupling is taken to zero, the non-triviality of
the result relies on the existence of the infinite set of fields
labeled by the fifth dimension. At the same time, in order to 
recover the full Lorentz invariance we have to restore the eight 
supercharges, by taking the limit $\mu \rightarrow 0$. Since the 
supergravity description is valid also for finite $\beta$, we 
will explore that case as well. 
As explained above, the specific
choice of supersymmetry breaking term allows us to explore only a
subspace of the full moduli space. It turns out that this is enough
to expose all the features predicted by the supergravity analysis.

\sect{The Superpotential and its distinguished points}

It is relatively easy to see from the superpotential (\ref{Wmin}) that
in the space of theories there exists a set of special points
distinguished by the integrality properties of $A=\frac{1}{4\pi^2}g_5^2
N m_0$, as predicted by the supergravity analysis. 

For ease of notation let us introduce the purely imaginary parameters
$z$ and  $\ttau$ as
\be
z=iz_2 \equiv i{\beta m_0 \over 2} ~~~~~~~~\ttau = i \ttau_2 
\equiv  i {2 \pi \beta \over g_5^2 N }~~.
\ee
With this notation the effective superpotential is written as
\bea
W_{\it eff} =- {N\mu \over 
\sin \, z}
{\theta_1 \left(z|\ttau \right) \over 
\theta'_1\left(0|\ttau \right)}
\label{w2p}
\eea
The decompactification limit which we are interested in,
$\beta\rightarrow\infty$, becomes now a double scaling limit
\begin{equation}
z_2,\ttau_2\rightarrow\infty~~~~~~~~{\rm with}~~~~ 
A=\frac{z_2}{\pi \ttau_2}={\rm fixed}~~.
\end{equation}

As it is usually the case with expressions involving theta functions,
different representations are useful for different purposes. To show
that $W_{\it eff}$ has special properties for integer 
$A$, the form (\ref{w2p}) is not particularly useful, since $A$ does
not appear explicitly. An $S$ modular transformation however casts it
in a more useful form:
\begin{equation}
W_{\it eff} =- {N\mu \over \ttau_2\sinh \, z_2} e^{\frac{1}{2}\beta
m_0 A} {\theta_1 \left(-A|-{1\over \ttau} \right) \over 
\theta'_1\left(0|\ttau \right)}
\end{equation}
Using now the product representation for the theta functions we
immediately find that
\be
W_{eff} = - {N\mu 2 \pi \beta \over g_5^2 N \sinh ~ {\beta m_0 \over 2}}  
e^{\beta m_0 A \over 2} \sin (\pi A) \prod_{n=1}^{\infty} 
(1+  {\sin^2 (\pi A) \over \sinh^2 ({n  g_5^2 N \over 2\beta}) })~~,
\label{product} 
\ee
Thus, for theories for which 
\begin{equation}
A\in \Z
\end{equation}
the effective superpotential vanishes for all compactification radii.
We should emphasize that the parameter $A=\frac{1}{4\pi^2}g_5^2Nm_0$ has
an intrinsic 5-dimensional meaning, as it does not depend on the 
compactification radius $\beta$ or on the tree level superpotential
breaking the 5 dimensional Lorentz invariance. This implies that, from
a 4 dimensional standpoint, the vanishing of the effective
superpotential relies entirely on the existence of the continuum of
fields labeled by the fifth coordinate. As we have seen in section 2,
the integer values of $A$  were also special in the supergravity dual
of this theory, being the only ones which allowed brane polarization.
We will come back to the relation between the two descriptions in the
next section where we discuss the effective superpotential for
the physical fields.

Although the product form of the superpotential exposes its zeroes,
it does not yield a well defined expansion for
$\beta\rightarrow\infty$ since all terms in the product
(\ref{product}) have contributions of equal magnitude. It turns out
that the series presentation of the theta
functions is more useful in the analysis of the decompactification
limit. From 
\begin{eqnarray}
\theta_1(z|\tau)=- i \sum_{n=-\infty}^\infty (-)^nq^{(n+1/2)^2}
e^{(2n+1) iz} 
\label{theta_sum}\\
\partial_z \theta_1(0|\tau)=2\sum_{n=0}^\infty (-)^n(2n+1)
q^{(n+1/2)^2}
\nonumber
\end{eqnarray}
with $q=\exp(i\pi\tau)= \exp(- \pi \tau_2)$, it is clear that only
relatively few terms 
will give the dominant contribution as $\beta\rightarrow \infty$. 
Using the explicit form of $q$ and completing the squares in the 
exponent in (\ref{theta_sum}) we find
\bea
\theta_1(z|\tau) =  -ie^{\pi \tau_2 A^2}
\sum_{n=-\infty}^\infty (-)^n e^{-\pi\tau_2(n+1/2-A)^2}~~.
\label{new_theta}
\eea

In the decompactification limit $\beta\rightarrow\infty$ the
derivative of the theta function in the denominator of $W_{\it eff}$
yields only minor contributions which combine with the exponent of the
prefactor in (\ref{new_theta}) into a perfect square. We find that
\be
W_{\it eff}=-2 N\mu \left(1+{\cal O}(e^{-2\pi\tau_2})\right)
e^{\pi \tau_2 (A-1/2)^2}
\sum_{n=-\infty}^\infty (-)^n e^{-\pi\tau_2(n+1/2-A)^2}
\label{newWeff}
\ee
From this form it is not hard to recover the result derived before
from the product presentation of theta functions. Thus, to evaluate
the sum in (\ref{newWeff}) it is convenient to split 
$A$ in its integer and fractional parts:
\be
A=k+\alpha~~~{\rm with}~~~k\in \Z~~~{\rm and}~~~\alpha\in[0,1)~~.
\ee
Shifting the summation index by $k$ it is easy to see that the
remaining sum in $W_{\it eff}$ is
\bea
\sum_{n=-\infty}^\infty (-)^n e^{-\pi\tau_2(n+1/2-A)^2}=
(-)^k\sum_{n=-\infty}^\infty (-)^n e^{-\pi\tau_2(n+1/2-\alpha)^2}
\eea
For any nonvanishing $\alpha$ this sum is dominated by the terms with
$n=0$ and $n=-1$. Since these two terms also lead to a vanishing $W_{\it
eff}$ for vanishing $\alpha$, it is justified to define the
effective superpotential in the decompactification limit to be
\be
 \lim_{\beta \rightarrow \infty} W_{\it eff}=(-1)^{k+1} 2 N\mu 
e^{\pi \tau_2 (A-1/2)^2} (e^{-\pi\tau_2(1/2-\alpha)^2} 
- e^{-\pi\tau_2(1/2+\alpha)^2})~~.
\label{imp}
\ee

This equation captures all phenomena occurring in the family of 5
dimensional ${\cal N}=1^*$ theories parametrized by $g_5^2$ and $m_0$ 
which are accessible from the submanifold of their Coulomb branch 
parametrized by the expectation value of $\cosh\beta\Phi_3$.
In the next section we will analyze in detail its consequences
and recover various features predicted by the supergravity analysis.

\sect{Reconstructing the supergravity/string theory predictions for
the moduli space}

As we have seen in the previous section, the DV approach provides a
description in terms of the superpotential (\ref{imp})
for the
5-dimensional ${\cal N}=1^*$ theory compactified on a circle of radius
$\beta$ with a nonvanishing ${\cal N}=2 \ra {\cal N}=1$ supersymmetry
breaking term $\mu$. This superpotential
diverges as  $\beta \rightarrow \infty$ and goes to zero when $\mu \ra
0$. Thus, at first glance it seems as the only quantities that one could
 compare with the supergravity result must be independent of
$\mu$. Nevertheless, it is also
possible to obtain meaningful quantities by taking double scaling
limits $\beta \rightarrow \infty$ with $\mu \ra 0$ such that some
physical quantities remain constant. 

In this section we use this strategy to show that the vacuum structure 
of the ${\cal N}=1^*$ theories changes drastically when $A\in \Z$, as 
predicted by the supergravity/string theory dual. The absence of 
polarization branches 
for $A \notin \Z$ and their presence when $A$ is an integer emerges
beautifully from the matrix model effective superpotential (\ref{imp}).

In \S6 we will present more evidence that the matrix
model shows  that new massless degrees of freedom appear when the
theory is on a polarization branch, and will compare these predictions
to the  supergravity results and to the results of a perturbative
computation. This 
hints toward the identification of the $A\in \Z$ theories with the RG
fixed points conjectured by Seiberg in \cite{seiberg}. These points
were further analyzed from field theoretic and geometric engineering
standpoints in \cite{5d}-\cite{DOKV}.

\subsection{The polarization branches}

Inspired by the supergravity description of the theory, it is natural 
to expect the polarization branch of the theory to be parameterized by
the radius of the D4/NS5 configuration, which corresponds in the field
theory  
to the expectation value of the sum of squares of transverse scalars. 
In our case, a measure of the radius of this configuration is 
\begin{equation}
\frac{\beta}{g_5^2} \rho^2 \equiv 
 \frac{\beta}{g_5^2}\langle {\rm Tr}[\Phi_1\Phi_2]\rangle = - {\partial W
\over \partial m_0} ,
\end{equation}
where the factor of $\beta$ on the left hand side is consistent with the
theory being the reduction of a 5 dimensional theory,  $m_0$ is
the 5 dimensional mass parameter\footnote{This factor is also
consistent with the normalization of the supersymmetry breaking
deformation, since in the limit of small $\beta$ this deformation
becomes an ordinary mass term, with the same normalization as above.}
and $\rho$ is a 4 dimensional chiral superfield. Note 
that the real part of $\rho^2$ is $\langle {\rm Tr}[X_5^2 - X_6^2 +
X_8^2-X_9^2]\rangle$.  
Thus, a positive $\rho^2$ corresponds to polarization in the 
$(5,\,8)$ plane, while a negative  $\langle \rho^2\rangle$ corresponds
to polarization in the $(6,\,9)$ plane.

The key observation is that the 5-dimensional superpotential 
is highly constrained by the 
8 supercharges, and can only have hypermultiplet 
mass terms or/and Yukawa couplings of the vector and hypermultiplets.
Hence, the value of the superpotential at the critical points 
vanishes. Therefore, only the 4-dimensional vacua in which the value
of the superpotential vanishes in the decompactification limit will
become vacua of the 5-dimensional theory.

It is now fairly easy to obtain the moduli space.
Since the effective superpotential depends on 
two parameters, $\mu$ and $\beta$, one can recover the 
decompactified $\N=1^*$ theory by taking a double scaling limit 
$\mu \ra 0,~ \beta \ra \infty$. Every double scaling limit
in which the superpotential vanishes will fix us somewhere on the
 moduli space.  This will allow us to map a part of this moduli space
which includes the polarization branch emanating from the origin of
the Coulomb branch.

The analysis naturally splits into two different regimes.
First, we consider the case $A\in \Z$, in which the effective
superpotential vanishes for arbitrary values of $\mu$ and $\beta$ and
then we discuss the regime in which the effective superpotential
vanishes only in the double scaling limit.

In the first case,  even though the effective superpotential
vanishes identically, its derivative is nonzero. More concretely, the
expectation value of the polarization modulus is given by
\begin{equation}
\langle \rho^2\rangle = 2 g_5^2 N \left[\mu e^{\frac{1}{2}\beta m_0
(A-1)}\right]
\label{A1}
\end{equation}
Except for the case $A=1$ which is not covered by this analysis, one
can take double scaling limits $\mu\rightarrow
0,~\beta\rightarrow\infty$ in which $\langle \rho^2\rangle$ 
can be fixed to arbitrary, continuous values. Thus, the 5 dimensional
theory has a continuum of vacua parameterized by the expectation value
of  $ \rho^2$, which are exactly the vacua dual to 
polarized configurations of arbitrary radius. Furthermore,
since the parameter $\mu $ is complex, the family of vacua is
two-dimensional, matching the supergravity prediction.

Before proceeding, let us stress that capturing the polarization vacua 
through the DV approach depends on the choice of the $\N=2 \ra \N=1$
breaking  deformation. If we had chosen another type of deformation,
we could have  conceivably  obtained only $\N=1$ vacua which lift to
vacua on the Coulomb  branch, and would have remained oblivious to the
existence of a polarization  branch. Even though the perturbation we
chose only maps the polarization branch for $A \ge 2$, 
this branch exists also for $A=1$, and  the investigation of the new 
degrees freedom we present in the next chapter is unchanged in this
case. However, the double scaling limit of (\ref{A1}) does not allow us
to access this  branch.

Let us now turn to the second case, $A\not\in\Z$. In this situation the
double scaling limit is more constrained, since to recover 
an $\N=2$ vacuum we need $W_{\it eff}\rightarrow 0$. Even if the 
algebra is more complicated than for integer $A$, extracting the 
large $\beta$ scaling is straightforward, if one notices that the 
mass parameter $m_0$ appears only in the combination $\exp(-\beta
m_0)$. Thus, it is clear 
that as far as the large $\beta$ scaling is concerned, the would-be
polarization modulus behaves as
\begin{equation}
\frac{\beta}{g_5^2}\langle \rho^2\rangle = {\partial W
\over \partial m_0} \sim \beta W_{\it eff}~~~~~~~~\Rightarrow~~~~
\langle \rho^2\rangle \sim W_{\it eff}~~.
\end{equation}
Since  $W_{\it eff}$ vanishes in the decompactification limit, 
the only vacuum allowed in the case of noninteger $A$ 
is the one with vanishing expectation value for the polarization
modulus. Hence, the field describing fluctuations in this
radial direction is massive.
This recovers the remaining part of the supergravity 
prediction about the structure of the space of ${\cal N}=1^*$
theories -- for non-integral $A$ the branes are forced to sit at the
origin of four of the five transverse directions.

With hindsight, it is clear that the construction described above is
the only possibility of reconstructing the superpotential while the
usual integrating in procedure of \cite{intriligator} (see
e.g. \cite{us} for its application in the matrix model context)
is doomed to fail since it is not possible to integrate in massless
fields. 

\sect{The new degrees of freedom of the fixed point theories}

In this section we explore the new degrees of freedom which appear 
in the theory at integer $A$. We first discuss the supergravity
predictions for the physics of the polarization branch and then
examine the way in which  the new degrees of freedom appear in the
DV description. Though we will not be able to explicitly identify some
of the
light fields on the gauge theory side, we will point out some
features which are consistent with their existence.
We conclude with a perturbative computation, along the
lines of \cite{ed, seiberg}, which points to the identification of the
$A\in \Z$ theories with the fixed points of the renormalization group
conjectured in \cite{seiberg}.

\subsection{Supergravity predictions}

In order to  understand the physics of these new vacua, it is useful to
examine the objects of the maximally  
supersymmetric  theory living on parallel D4 branes. This theory
has  W-bosons and ``W-strings'', 
corresponding to F-strings and D2 
branes stretched between the D4 branes. The theory also has a massive
monopole (instanton), which is a D0 brane. When the theory is on the
Coulomb branch, the W-bosons and W-strings acquire  mass by the Higgs
mechanism. The mass of the monopole is unchanged.

The objects in the perturbed theory are the same, but their properties
change drastically on the polarization  branch. Since near an NS5 brane 
the dilaton diverges, the masses of the D0 and D2 brane
vanish and thus the theory has massless monopoles and massless
``W-strings''.  The mass of the W-boson is proportional to the
distance between the would-be location of D4 branes, which is
proportional to the radius of the  polarized configuration. Thus, at a
generic point on the polarization branch, the W-bosons are
massive. However, as one approaches the origin of the  
polarization branch along the polarization branch, the mass of 
these W-bosons also goes to zero. 

Hence, the W-bosons and W-strings, which couple electrically and
magnetically  to the gauge field on the brane, are massless; the
other objects in  the theory are massless as well. This suggests that
approaching the origin  from the polarization branch, brings us to a
nontrivial fixed point of the  renormalization
group flow that bears certain similarities to Argyres-Douglas type
theories in four dimensions \cite{ad} \footnote{It is important to note
that approaching the origin of the polarization branch from the
Coulomb branch does not  give the same point. Indeed, the monopoles
(D0 branes) are never massless on the Coulomb branch. Moreover, when
the branes are on the Coulomb branch  supergravity is never weakly
coupled in their vicinity. Therefore, the  IR physics is described by
the field theory, and we cannot use the string  descriptions of
W-bosons and W-strings to conclude they are massless.}. 
The fact that all the objects
of this  theory become massless also suggests that, much like in the
Argyres-Douglas  case, the theory becomes superconformal\footnote{It
is however not obvious how to compute the $\beta$ function and show
that it vanishes.}.

It is perhaps an appropriate moment to return to the observation made
in \S2 that the area of the horizon of the near extremal solution
(\ref{bh}) is independent of the deformation parameter $B$ for fixed
temperature. This implies that, for fixed temperature, the entropy of
the field theory is independent on the mass of the
hypermultiplet. This might seem to contradict the fact
that by varying $m$ it is possible to reach
points where the theory has extra light degrees of freedom.  

This puzzle is solved by observing that these light degrees of 
freedom are nonperturbative, and thus only visible in supergravity. 
As the solution is taken off extremality, only the
origin of the Coulomb branch survives  as a possible vacuum of the
theory. Both the Coulomb branch and the polarization branch are lifted 
by the temperature deformation. However, as pointed out above, the
origin does  
not belong to the
polarization branch and thus the extra massless degrees of freedom are
not part of the field theory excitation spectrum.

\subsection{The effective coupling}

As we have seen in the previous section, to recover the 5-dimensional 
theory we are interested in, we must take the double scaling limit
$\beta \ra \infty,  ~\mu \ra 0$ with $W_{\it eff}\rightarrow
0$. However, some of the quantities of this theory are independent  
of $\mu$, which makes them easy to obtain and study even
before taking this limit.

One such quantity is the effective gauge coupling. 
The relation between 4 dimensional ${\cal N}=1$ gauge theories and
special geometry \cite{dv123} implies that this
coupling is independent of the overall scale of the tree level
superpotential. This property survives also in the present case, as it
can be easily seen from equation (\ref{ww}). Thus, if
\be
2\pi i ~\tau_{eff} = {\partial ^2 F \over \partial S^2}
\label{teff}
\ee
is well-defined, then it is also independent of $\mu$. Explicitly
computing the effective coupling requires one to (at least in
principle) be able to invert the equation $S=S(t)$. This is only the 
case if $h''(t) \neq 0$ since
\begin{equation}
\frac{\partial^2 F}{\partial t ~\partial S} = t h''(t)~~~~~~~~
\frac{\partial S}{\partial t} = \frac{1}{2\pi i} h''(t)~~.
\end{equation}
Thus, in the case of the nondegenerate solutions of (\ref{vacua}) 
the effective coupling is constant and given by the 't Hooft coupling
\begin{equation}
\tau_{\it eff} = \frac{\tau}{N}~~.
\label{hol}
\end{equation}

Before we proceed to the degenerate case, let us note that this seems
quite different from the effective coupling computed in \cite{nekrasov}, or 
the one we will compute in \S6.2 using perturbative arguments.

The difference arises from the fact that the the supersymmetry
breaking deformation (\ref{breaking}) selects a vacuum which is at
finite distance from the origin of the Coulomb branch. 
Even though the superpotential
(\ref{Wmin}) does not allow us to explicitly compute $ \langle
\,{\rm Tr}[\Phi_3^{2 k}]\,\rangle$ for all $k$, we can still estimate 
some of them in the decompactification limit. 
From equation (\ref{imp})\footnote{Using $
\langle \,{\rm Tr} (\cosh (\beta \Phi_3)-1)\,\rangle = {\partial
W_{\it eff} \over \partial \mu}$} we find
\begin{eqnarray}
\langle \,{\rm Tr}[\Phi_3^2]\,\rangle ^{1\over 2} ~& \sim &~ 2 m A 
~~~~~~~{\rm for}~~~~~~~A \not\in { \Z}~~,
\label{estimates}
\end{eqnarray}
and similar expressions for the other invariants. 
Thus, although we have started from a configuration where $\Phi_3$ is 
zero (which was reflected in the choice for the matrix model saddle
point), the multilinears in this field acquire 
nontrivial expectation values at the quantum level.

Equation (\ref{hol}) is also consistent with
the supergravity description of the ${\cal N}=1^*$ theories. Indeed, 
we can show that by reducing to 4 dimensions the vacua 
selected by the supersymmetry breaking
term in (\ref{breaking}), we find (up to possible T
dualities) a background where the dilaton near the branes is 
independent of $m$. 

From (\ref{estimates}) it is easy to see that the size of the D-brane 
distribution dual to the selected Coulomb branch configuration is of
 order $m A \sim m^2 g_s N$. For this configuration, supergravity is a valid description for $m\ll r\ll \frac{1}{m}$, which for a fixed $A$ 
translates into
\begin{equation}
\frac{1}{g_5^2 N}\,\ll \,\, r\,\,\ll \,g_5^2 N~~.
\end{equation}
i.e. the supergravity description is valid almost in
the entire space. 

For a distribution of this size, close enough to the branes,
 the two terms defining the dilaton
in (\ref{bg}) are of the same size. Therefore, near the branes 
the dilaton behaves as 
\begin{equation}
\frac{e^{4\phi/3}}{g_s^{4/3}}=\Lambda
=Z^{-1/3}\left(1+\alpha\right)~,
\end{equation}
where $\alpha$ is a constant independent of $m$.
The metric coefficients have a similar behavior. In particular, the 
coefficient corresponding to the new compactified direction is
proportional to $Z^{-1/2}$, up to a constant independent of $m$. 
These are (up to some irrelevant numerical
coefficients) precisely the characteristics of an unperturbed D4 brane
solution. It is then clear that by T-dualizing along the circle
the dilaton becomes $m$-independent, as predicted by
the DV analysis (\ref{hol}).

Let us now return to the case of the degenerate solutions in
(\ref{vacua}). Since the relation between $S$ and $t$ cannot
be inverted, the effective coupling $\tau_{\it eff}$ is
undetermined. Moreover, in the decompactification
limit,   
the nondegenerate vacua of theories with $A \in \Z$ asymptote to
degenerate vacua\footnote{One can easily see this by noticing that the
points where $h(t)=0$  and $h''(t)=0$ are separated by $\delta t \sim
{1 \over \beta}$.}. 

Therefore, for integer $A$, $\tau_{\it eff}$ is undetermined by
(\ref{teff}). These 
features suggest a breakdown of the low energy effective theory which
can occur due to  the appearance of new light degrees of freedom.
This interpretation is consistent with the supergravity analysis
discussed  
in the previous subsection.

The effective coupling $\tau_{\it eff}$ is related to the expectation
value of the bulk dilaton, which varies as one changes the
polarization radius. Therefore the  value of $\tau_{\it eff}$, though
undetermined by gauge theory, can be used to parametrize (at least
locally) the moduli space of vacua. The degeneracy discussed above is
therefore 
a reflection of the  fact that to one classical vacuum corresponds a
continuum of quantum vacua, indistinguishable in the particular matrix
model used here. 

To summarize, when  $A\not\in \Z$ the effective coupling is 
well-defined and proportional to the bare coupling at all points on
the Coulomb branch selected by the $\N=2 \ra \N=1$  deformation term
(\ref{breaking}). However, when $A\in \Z$, $\tau_{\it eff}$  
is undetermined and the expression of the glueball superfield $S$ is
degenerate, suggesting a breakdown of the low energy effective
description and  the appearance of new light degrees of freedom.

\subsection{Perturbative breakdown}

Some of the special points singled out both by the supergravity
and the DV description of the 5 dimensional ${\cal N}=1^*$ theory 
can also be identified perturbatively, using the techniques 
of \cite{seiberg} and \cite{ed}. It is not hard to compute the 
effective coupling of 5-dimensional theories since, 
as argued in \cite{seiberg}, it is only corrected at one loop. These
corrections  lead to the vanishing of the effective $\tau$ parameter,
and imply the  existence of a regime of ultra-strong coupling. This
will occur at the first nontrivial points in the series of theories
obeying 
\begin{equation}
A\in \Z
\label{integerA}
\end{equation}

Since the effective coupling constant is related to a 5 dimensional
Chern-Simons term, it receives 
a finite shift at 1 loop. According to \cite{seiberg}, \cite{ed}, in a
theory containing hypermultiplets of masses $m_i$, this
correction has the general form
\begin{equation}
\frac{1}{g_{\it eff}^2} = \frac{1}{g_5^2} + a\phi - \sum_i b_i |\phi
    - m_i| 
\end{equation}
where $\phi$ is the classical expectation value of the scalar in the
vector multiplet. The first term on the right hand side is 
the bare gauge coupling , and the following ones are 
due to hypermultiplet interactions. The coefficients $b_i$ depend on 
the representation of the $i^{th}$
hypermultiplet. It is easy to see that these terms arise at the
1-loop level from the self-energy diagram of the vector multiplet, and 
therefore their coefficients are equal to  ${\rm Tr}_{R_i}[T^a T^b]$
in the appropriate representation $R_i$, up to a universal coefficient. 

To match the assumptions which led to (\ref{Wmin}), we are interested
in the configuration at the origin  of the Coulomb branch (where the
gauge group is classically unbroken) which corresponds to $\phi=0$. In
this case only the hypermultiplets contribute;  
furthermore, since there exists only one hypermultiplet (of mass $m$, 
transforming in the adjoint representation), the 1 loop effective
coupling is given by 
\begin{equation}
\frac{1}{g_{\it eff}^2} = \frac{1}{g_5^2} - {1 \over 4 \pi^2} N |m|~~.
\label{s-p}
\end{equation}
Consequently, the effective 5-dimensional coupling vanishes for
\begin{equation}
 g_5^2 N |m|=4 \pi^2
\end{equation}
which is the first case in the series (\ref{integerA}).

In \cite{seiberg} such points were given the interpretation of 
fixed points of the renormalization group, exhibiting enhanced
symmetries. From the supergravity analysis it is 
certainly the case that a number of mutually nonlocal massless states
appear at 
these points. From that analysis, it is however not clear what is the
enhanced global symmetry. Clearly, it
would be interesting to investigate this further.

\sect{Conclusions and Future Directions}

We have compared the supergravity/string theory and the matrix model 
descriptions of 
the $\N=1^*$ theory in 5 dimensions, and have found that the unusual 
properties displayed by these theories for $\frac{1}{4\pi^2}{g_5^2 N
m} \in \Z$  are captured by both approaches. 

The existence of a nonperturbative branch of the moduli space, 
which in the supergravity corresponds to the possibility of polarizing
D4 branes into  NS5 branes, emerges in the matrix model from
certain properties of Jacobi $\theta$ functions. 
The theories along the polarization branch posses new light 
degrees of freedom.
At the supergravity level they are the polarization modulus as well as
D0 and D2 branes which become massless near the polarized D4 branes.
In the matrix model description, these new  degrees of freedom
manifest themselves through the breakdown of the effective description
at the special points  $\frac{1}{4\pi^2} {g_5^2 N m} \in \Z$. 

We have also compared the matrix model description of 
the 5-dimensional ${\cal N}=1^*$ theory to the integrable models-based
Seiberg-Witten-like description \cite{nekrasov}, as well as 
with perturbative arguments along the lines of 
\cite{seiberg,ed}. These 4 descriptions agree with  each other in
the ranges of mutual validity.

The supergravity description allowed us to prove a conjecture of
Nekrasov that a certain shift of the hypermultiplet mass is a symmetry
of this theory. We have also computed the entropy of this theory,  
and found that at finite temperature it is independent of $m$, which
is again consistent with a shift in $m$ being a symmetry of the
theory. The picture which emerges is that the spectrum of the theory
is reshuffled under this transformation.

Perhaps the most intriguing question raised by our analysis is {\em
why} are the  theories with $A \in \Z$ special. Except the possibility 
for brane polarization, the supergravity duals show no special
features at these particular points. On the gauge theory side one
might contemplate a 6-dimensional interpretation\footnote{A loosely
similar example with these characteristics was discussed in
\cite{rt94} where it was shown that the partition function of string
theory on a Melvin background exhibits certain equally-spaced
zeroes. The analogous analysis  
in the case at hand is unfortunately prohibitive.}. The lift to 
six dimensions of the
$\N=1^*$ theory  is the (2,0) theory perturbed with a vector potential
which breaks 6-dimensional Lorentz invariance. This theory and its
gravity dual have been studied in \cite{diana},  
and no indication of any special points has emerged. 
It is possible that a DLCQ description of these theories might
reveal aspects of the special physics associated with these
points. However, there is no straightforward 6 dimensional (2,0)
explanation of {\em why} these points are special. 
A  source of the difficulties can be traced to the fact that
the 5 dimensional coupling emerges only in the compactification
process and has no interpretation in the strict six dimensional limit.

The setup discussed in this paper can be generalized by adding
of other branes, orbifolds and orientifolds, to include other
gauge groups and matter representations.
The perturbative arguments described in \S6.3 apply 
rather straightforwardly to these theories and suggest that, at least
in some cases, they should also have an infinite set of points where
exceptional physics emerges. This could also be a hint that such theories 
have a shift symmetry similar to the one studied here, while the arguments of
\cite{seiberg} would suggest that these points are fixed under RG
flow. 

These two possibilities could be ascertained in three ways: by
providing a supergravity dual to these theories (which is probably
straightforward for the orbifold/orientifold   
case) by solving the quantum mechanics with $SO(N),Sp(N)$ gauge groups
and different matter representations or by solving the integrable
system analogous to the one in \cite{nekrasov}. It is however not {\it
a priori} clear which approach is the most profitable one.

It would also be interesting to solve the matrix quantum mechanics  
for a generic $\N=2 \ra \N=1$ deformation, and thus map the full
moduli space of the 5-dimensional $\N=1^*$ theory. This would allow
one to confirm the supergravity prediction that the Coulomb branch
intersects the polarization branch at all points of unbroken  
gauge symmetry, and to study the way one moves between these two
branches. 

Another interesting direction to explore is the limiting
behavior of the theories along the polarization branch as the radius of
polarization ($\langle {\rm Tr}[\Phi_1\Phi_2]\rangle $)  
goes to infinity. In this limit the circular D4-NS5 brane 
configuration decompactifies. Although ten dimensional 
supergravity might not valid in this regime  
(unless one also scales $N$ with the radius), the field theory analysis
indicates that such a limit makes sense. It would be interesting to
study the theory in this limit and to determine whether in this case  
it is possible to find a 6-dimensional explanation of the exceptional
5-dimensional physics. 

Many of the arguments which lead to the phenomena we have discovered 
for the $\N=1^*$ theory in 5-dimensions seem to generalize 
to other 5-dimensional theories. This may lead one to believe that
nonperturbative branches of the moduli space and mass shift
periodicities are generic features of field theories in 5 dimensions.
Understanding whether this is indeed the case, and if so why, is an
important open question. 

\vskip 5mm

\noindent
{\bf Acknowledgments}

\vskip 3mm
We would like to thank Cumrun Vafa, Per Kraus, Joe Polchinski, Eric
d'Hoker, Alex Buchel, Robert Dijkgraaf, Sergei Gukov, Nick Warner, 
Gary Horowitz, Michael Gutperle, Oliver DeWolfe, Steve Giddings 
and Radu Tatar for
useful comments and discussions. 
The work of I.B. was supported by the NSF under Grants No. PHY00-99590 
and PHY01-40151. The work of R.R. was supported in part by the DOE under 
Grant No. 91ER40618 and in part by the NSF under Grant No. PHY00-98395.


\newpage

\begin{thebibliography}{10}
\baselineskip=15pt

\bibitem{imsy}{
N.~Itzhaki, J.~M.~Maldacena, J.~Sonnenschein and S.~Yankielowicz,
Phys.\ Rev.\ D {\bf 58}, 046004 (1998)
[arXiv:hep-th/9802042].
}

\bibitem{rc}{
I.~Bena and C.~Ciocarlie,
arXiv:hep-th/0212252.
}

\bibitem{myers}
R.~C.~Myers,
JHEP {\bf 9912}, 022 (1999)
[arXiv:hep-th/9910053].

\bibitem{n=1*}{
N.~Dorey,
JHEP {\bf 9907}, 021 (1999)
[arXiv:hep-th/9906011].
}

\bibitem{dv4}{
R.~Dijkgraaf and C.~Vafa,
arXiv:hep-th/0302011.
}

\bibitem{dv123} 
R.~Dijkgraaf and C.~Vafa,
Nucl.\ Phys.\ B {\bf 644}, 3 (2002), hep-th/0206255.

R.~Dijkgraaf and C.~Vafa,
Nucl.\ Phys.\ B {\bf 644}, 21 (2002), hep-th/0207106.

R.~Dijkgraaf and C.~Vafa,
hep-th/0208048.

\bibitem{holo}{
T.~J.~Hollowood,
JHEP {\bf 0303}, 039 (2003)
[arXiv:hep-th/0302165].
}

\bibitem{nekrasov} {
N.~Nekrasov,
Nucl.\ Phys.\ B {\bf 531}, 323 (1998)
[arXiv:hep-th/9609219].
}

\bibitem{malda}{
J.~M.~Maldacena,
Adv.\ Theor.\ Math.\ Phys.\  {\bf 2}, 231 (1998)
[Int.\ J.\ Theor.\ Phys.\  {\bf 38}, 1113 (1999)]
[arXiv:hep-th/9711200].
}

\bibitem{bdhm-bklt}{
T.~Banks, M.~R.~Douglas, G.~T.~Horowitz and E.~J.~Martinec,
arXiv:hep-th/9808016.
}

{
V.~Balasubramanian, P.~Kraus, A.~E.~Lawrence and S.~P.~Trivedi,
Phys.\ Rev.\ D {\bf 59}, 104021 (1999)
[arXiv:hep-th/9808017].
}

\bibitem{ps}{
J.~Polchinski and M.~J.~Strassler,
arXiv:hep-th/0003136.
}

\bibitem{gs}{
M.~Gutperle and A.~Strominger,
JHEP {\bf 0106}, 035 (2001)
[arXiv:hep-th/0104136].
}

\bibitem{fig}{
J.~Figueroa-O'Farrill and J.~Simon,
arXiv:hep-th/0208107.
}

\bibitem{maldacena}
J.~M.~Maldacena and C.~Nunez,
Int.\ J.\ Mod.\ Phys.\ A {\bf 16} (2001) 822
[arXiv:hep-th/0007018].

\bibitem{guster} S.~S.~Gubser,
Adv.\ Theor.\ Math.\ Phys.\  {\bf 4} (2002) 679
[arXiv:hep-th/0002160].

\bibitem{rt} 
J.~G.~Russo and A.~A.~Tseytlin,
Nucl.\ Phys.\ B {\bf 461}, 131 (1996)
[arXiv:hep-th/9508068].

\bibitem{siegel} 
N.~Marcus, A.~Sagnotti and W.~Siegel,
Nucl.\ Phys.\ B {\bf 224} (1983) 159.


\bibitem{arkani-hamed} 
N.~Arkani-Hamed, T.~Gregoire and J.~Wacker,
JHEP {\bf 0203} (2002) 055
[arXiv:hep-th/0101233].

\bibitem{dorey}
I.~K.~Kostov,
Nucl.\ Phys.\ B {\bf 575}, 513 (2000)
[arXiv:hep-th/9911023].

N.~Dorey, T.~J.~Hollowood and S.~P.~Kumar,
JHEP {\bf 0212}, 003 (2002)
[arXiv:hep-th/0210239].

\bibitem{leigh} 
R.~G.~Leigh and M.~J.~Strassler,
Nucl.\ Phys.\ B {\bf 447}, 95 (1995)
[arXiv:hep-th/9503121].

\bibitem{seiberg}{
N.~Seiberg,
Phys.\ Lett.\ B {\bf 388}, 753 (1996)
[arXiv:hep-th/9608111].
}


\bibitem{lawrence-nekrasov}{
A.~E.~Lawrence and N.~Nekrasov,
Nucl.\ Phys.\ B {\bf 513}, 239 (1998)
[arXiv:hep-th/9706025].
}

\bibitem{ed}{
E.~Witten,
Nucl.\ Phys.\ B {\bf 471}, 195 (1996)
[arXiv:hep-th/9603150].
}

\bibitem{5d}{
O.~Aharony, A.~Hanany and B.~Kol,
JHEP {\bf 9801}, 002 (1998)
[arXiv:hep-th/9710116].
}
{
\bibitem{INMS}K.~A.~Intriligator, D.~R.~Morrison and N.~Seiberg,
Nucl.\ Phys.\ B {\bf 497}, 56 (1997)
[arXiv:hep-th/9702198].
}
{
\bibitem{MOSE}D.~R.~Morrison and N.~Seiberg,
Nucl.\ Phys.\ B {\bf 483}, 229 (1997)
[arXiv:hep-th/9609070].
}
{
\bibitem{DOKV}M.~R.~Douglas, S.~Katz and C.~Vafa,
Nucl.\ Phys.\ B {\bf 497}, 155 (1997)
[arXiv:hep-th/9609071].
}

\bibitem{intriligator}{
K.~A.~Intriligator,
Phys.\ Lett.\ B {\bf 336}, 409 (1994)
[arXiv:hep-th/9407106].
}

\bibitem{us}{
I.~Bena, R.~Roiban and R.~Tatar,
arXiv:hep-th/0211271.
}

\bibitem{ad} 
P.~C.~Argyres and M.~R.~Douglas,
Nucl.\ Phys.\ B {\bf 448}, 93 (1995)
[arXiv:hep-th/9505062].

\bibitem{diana}
I.~Bena and D.~Vaman,
JHEP {\bf 0111}, 032 (2001)
[arXiv:hep-th/0101064].

\bibitem{rt94}
J.~G.~Russo and A.~A.~Tseytlin,
Nucl.\ Phys.\ B {\bf 448}, 293 (1995)
[arXiv:hep-th/9411099].

\end{thebibliography}
\end{document}